\input{epsf}
\documentclass[journal,twoside]{IEEEtran}

\usepackage{graphicx}  

\usepackage{amsmath}   
\begin{document}
%
\title{On the Performance of Selective Relaying}
\title{On the Performance of Selection Relaying}
%
%
\author{Abdulkareem~Adinoyi$^1$, 
Yijia Fan$^2$,
         Halim~Yanikomeroglu$^1$, and H.~Vincent~Poor$^2$ \\
         \small $^1$Broadband Communications and Wireless Systems
         (BCWS) Centre \\
        \small  Dept. of Systems and Computer Engineering, \\
        \small  Carleton University, Ottawa, Canada \\
        \small $^2$Department of Electrical Engineering\\
         \small Princeton University, Princeton, NJ, USA
\thanks{The work was conducted in part at the Broadband Communications and
Wireless Systems (BCWS) Centre, Department of Systems and Computer
Engineering, Carleton University, Ottawa, Canada (e-mail:
adinoyi@sce.carleton.ca, yijiafan@princeton.edu,
halim@sce.carleton.ca, poor@princeton.edu). Partial support was
provided by the U.S. National Science Foundation under Grants
ANI-03-38807 and CNS-06-25637.} }
\maketitle
\begin{abstract}
Interest in selection relaying is growing. The recent developments in this area have
 largely focused on information theoretic analyses such as outage performance. Some of these analyses are
 accurate only at high SNR regimes. In this paper error rate analyses that are sufficiently
 accurate over
a wide range of SNR regimes are provided. The motivations for this
work are that practical systems operate at far lower SNR values than
those supported by the high SNR analysis.  To enable designers to make informed decisions regarding
 network design and deployment, it is
imperative  that system performance is evaluated with a reasonable
degree of accuracy over  practical SNR regimes. Simulations have been used to corroborate the
analytical results, as close agreement between the two is observed.
\end{abstract}

%
%
%
%
%

\section{Introduction and Motivation}
Selection diversity is a fundamental technique that can be
transferred over from traditional multiple antenna systems to
cooperative relaying systems. As add-on features to network relays,
cooperative techniques should not impose strict limitations or
require sophisticated hardware. This view of cooperation in relay
networks is informed by the fact that future wireless communication
standards will be relay-enhanced. The activities in IEEE 802.16 j/m
attest to the vital role relays would  be playing in future broadband
communication networks. Basically, the aim of deploying these relays
is to break the link between two communicating nodes into smaller
ones (links) with the objective of enabling high data rates and
coverage extension. Our view on relay cooperation is that cooperation
comes as an  add-on feature, which attempts to  extract additional
benefits on top of this primary objective.

Interest in various forms of cooperative schemes has been growing
steadily since the seminal work in cooperation diversity
in~\cite{laneman2004} and~\cite{sendonaris2003a}.  In these works,
it is shown that re-examining the manner in which network nodes
relate to each other can provide a cost-effective way of harnessing
the advantages (such as diversity gains) of multi-antenna systems
without  necessarily putting these antennas in one location. This
new paradigm has been known in the literature as cooperative
relaying or user cooperation diversity schemes, and it is
particularly attractive for small-size, antenna-limited wireless
devices.

However, in parallel cooperative relaying where each relay is
equipped with a single  antenna, the potential diversity gains in
the network will be destroyed if any of the relays attempts to fully
decode its received signal; the performance will be  bounded by the
single antenna system~\cite{Hua2003}. Selective relaying techniques
are often employed to overcome this problem. In particular the work
in~\cite{furuzan2007} discusses selective relaying in the context of
single relay that has knowledge of the channel states, helping a source. The present
paper considers selection relaying that involves upper layers of the
communication protocol. In other words, it is an overlay technique
on the routing mechanisms.


The selection relaying schemes analyzed in this work are closely related to
those in~\cite{beres2008,aggelos2006} with the following differences:
Here, we provide analysis for the error rate while these earlier
works focused on information theoretic analyses. The analysis
performed in~\cite{beres2008} is for the large SNR regime.  
The   implications of always combining the relay-destination path
with the source-destination path are not apparent given the
conclusions on parallel relays found in~\cite{Hua2003}. Should  the
selected relay always  cooperate with the source?  Our analyses show
that combining the relay path with the source-destination path also
provides full diversity. In addition, we show that selecting one
path while considering the direct path as a virtual relay path also
provides full  diversity.

Finally, complementary to the outage analyses in~\cite{beres2008}
and \cite{aggelos2006}, we provide expressions for evaluating the
outage probability and  capacity of the selection relaying schemes.
Our outage probability expressions are exact in both the low and
high SNR regimes. The motivation for the low or medium SNR analysis
is the following. It is noted that large SNR analysis has
theoretical merits; however, practical systems often  operate at
lower SNR values. Thus, it is desirable to be able to evaluate
system performance to a reasonable degree of accuracy in the low SNR
region for the purpose of network design and
 deployments.

\section{System Model}
\begin{figure}
\centering
 \includegraphics[scale=0.55]{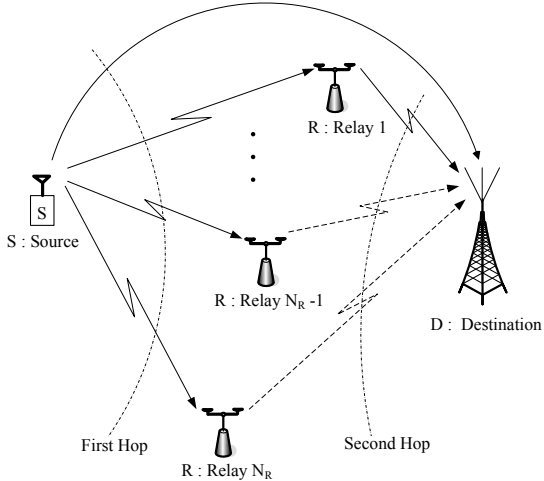}
 \vspace{-8cm}
  \caption{A multi-relay network with selection relaying.}\label{systemmodel}
\end{figure}

The system investigated in this paper is shown in
Fig.~\ref{systemmodel}. The source,  destination, and relays are
denoted as S, D, and R$_r \in \{1, \cdots, N_R\}$, respectively.
Each node is equipped  with a single antenna. The best relay (chosen
by some routing scheme) assists S-D communication. A block fading
channel model is assumed, in which the channel does not change in
the block. However, different blocks experience different channel
samples that  are
 independently and identically distributed
Rayleigh random variables. $\gamma_{0}$  denotes the instantaneous
SNR of the S-D link, while  $\gamma_{1} $ represents the minimum SNR
of the S-R and R-D links for the best relay. The Rayleigh fading
model implies that $\gamma_0$ is an exponential random variable, and
we denote its expected value by $\bar{\gamma_0}$. The statistics of
$\gamma_1$ are discussed below.

\vspace{-0.3cm}
In the half-duplex two-hop protocol,  S may or may not be able to
communicate directly with the destination.  The inability of the
source to do so  may be due to heavy  shadowing. During the first
hop, the source transmits while all the $N_R$ listen. During the
second hop, the best relay based on a metric that depends on the S-R
and R-D channels is selected to forward to the destination. The
protocol assumes that the network has a mechanism to select this
best relay.
To that effect we  state that a number of algorithms have been
proposed in the literature for performing this
task~\cite{aggelos2006}. Thus, the objective of this paper is not to
revisit path selection algorithms, but to focus  on the analysis of
selection relaying schemes.  In our analysis, it will be assumed
that the destination uses the signal received through the relay
according to selection relaying, SR (the S-D path is not usable at
the destination), selection cooperative relaying, SCR (the S-D link
is used at the destination) and all-path selection relaying, ASR
(the S-D link is among the path selection mechanism).

\vspace{-0.3cm}
As mentioned above implementation issues relating to joint selection
relaying schemes have been previously examined; it has been observed
that selecting the best relay can be performed at a radio network
controller, which comes at the expense of increased system overhead.
To reduce the overhead and system complexity,~\cite{beres2008}
considered a simplified relay selection scheme based only on the R-D
channel.

\section{Performance Analysis}

\subsection{Probability of Bit Error Calculations: Selection Cooperative  Relaying (SCR)}
The relay selection, or more accurately the  path selection, is
based on backward and forward channels and is performed jointly.
This means  that we select the relay $r^*$  with $\gamma = \max_r
\min (\gamma_{s,r}, \gamma_{r,d})$, where $\gamma_{s,r}$ and
$\gamma_{r,d}$ are the instantaneous SNRs of the S-R and R-D links,
respectively. The signal transmitted by this selected relay is
combined with the direct path signal using maximal ratio combining
technique. Therefore,  the combined SNR at the destination  is the
sum of the two SNRs.  Note that the destination uses the weaker of
the first and second hop of the selected relay. In terms of
capacity, this weaker link constitutes the bottle-neck as far as the
end-to-end performance is concerned.

Using the above notations, after maximal ratio combining, the SNR is
given as $\beta = \gamma_0 + \gamma_1$. Due to the independence of
$\gamma_0$ and $\gamma _1$, the (probability density function (PDF)
of $\beta$ can be obtained through the convolution of the PDFs of
$\gamma_1$ and $\gamma_0$, i.e.,
 \begin{eqnarray}\label{pdf_conv}
        p(\beta) &=& \int_{0}^{\beta}p_{\gamma_0}(\tau) \, p_{\gamma_1}(\beta - \tau) \, d\tau.
 \end{eqnarray}
As noted earlier, the selection of the best relay requires order statistics. 
The first step is to obtain the weaker link between the first hop
and second hop of each relay node. These  weak links are ordered and
the one with the largest SNR is selected as the candidate relay to
perform detection and forwarding to the
 destination. Given the PDF $f(\gamma)$ and CDF $F(\gamma)$ of the underlying Rayleigh distributed random variable, the PDF of such ordered random variables can be obtained~\cite{CohenBook1991}~\cite{Dai2007} as
$p(\gamma) = 2 N_R \, f(\gamma) \left[1 - F(\gamma))\right]\, \left[2F(\gamma) - (F(\gamma))^2)\right]^{N_R -1}.$
Since  $f(\gamma) = \frac{1}{\bar{\gamma}}
\exp\left(\frac{-\gamma}{\bar{\gamma}}\right) $ and $F(\gamma) = 1 -
\exp\left(\frac{-\gamma}{\bar{\gamma}}\right)$, the PDF of
$\gamma_1$ can be obtained as,
\begin{equation}
p(\gamma_1) =  N_R \, \frac{\exp(-\frac{\gamma_1}{\bar{\gamma_1}/2})}{\bar{\gamma_1}/2} \bigg(1 - \exp \bigg(-\frac{\gamma_1}{\bar{\gamma_1}/2}\bigg)\bigg)^{N_R -1},
\label{pdf2}
\end{equation}
and through    binomial expansion, we further can write
%
\begin{equation}
p(\gamma_1) = \sum_{i=1}^{N_R} \left(-1\right)^{i-1} \, \left( \begin{array}{c} N_R \\ i  \end{array}\right) \frac{2\,i}{\bar{\gamma_1}} \,
\exp \left( - i \, \frac{2\gamma_1}{\bar{\gamma_1}}\right).
\label{pdf_SR}
\end{equation}

Using  (\ref{pdf_SR}) and the  PDF of $\gamma_0$ (i.e.,  $\frac{1}{\bar{\gamma}_0} \exp (-\frac{\gamma_0}{\bar{\gamma}_0})$), we have
 \begin{eqnarray}
 p(\beta)&=&   \int_{0}^{\beta} \sum_{i=1}^{N_R} \left(-1\right)^{i-1} \,  \bigg( \begin{array}{c} N_R  \\ i  \end{array}\bigg) \frac{2\,i}{\bar{\gamma_1} \bar{\gamma_0} } \nonumber
\\ &\times&
\exp \left( -  \frac{i \, \tau}{\bar{\gamma_1}/2}\right) \exp \left(- \frac{\beta  - \tau}{\bar{\gamma}_0}  \right) d\tau.
\label{pdf5}
\end{eqnarray}
By interchanging the integral and summation, (\ref{pdf5}) can be expressed as
 \begin{eqnarray}
 p(\beta)&=&    \sum_{i=1}^{N_R} \left(-1\right)^{i-1} \,  \bigg( \begin{array}{c} N_R \\ i  \end{array}\bigg) \frac{2\,i}{\bar{\gamma_1} \bar{\gamma_0} } \nonumber
\\ &\times&
\exp \left(-\frac{\beta}{\bar{\gamma}_0} \right) \int_{0}^{\beta} \exp \left( -  \bigg[ \frac{i \, \tau}{\bar{\gamma_1}/2} - \frac{\tau}{\bar{\gamma}_0}  \bigg] \right)  d\tau.
\label{pdf6}
\end{eqnarray}
Finally,
 \begin{eqnarray}
 p(\beta) &=&    \sum_{i=1}^{N_R} \left(-1\right)^{i-1} \, \bigg( \begin{array}{c} N_R  \\ i  \end{array}\bigg)
 \frac{2\,i}{2 i \bar{\gamma_0} - \bar{\gamma_1} } \nonumber \\
&\times& \left(\exp \left[-\frac{\beta}{\bar{\gamma}_0} \right] - \exp\bigg[- \frac{2\,i \, \beta}{\bar{\gamma_1}} \bigg]\right).
\label{pdf_SCR}
\end{eqnarray}
The PDF obtained in (\ref{pdf_SCR}) can be employed for  evaluating
the error performance of this relaying scheme with any modulation
technique. However, we will demonstrate the evaluation with binary
phase shift keying (BPSK) as follows:
\begin{eqnarray}
BER_{scr} &=& \frac{1}{2} \int_0^{\infty} \mbox{erfc}\left(\sqrt{\beta}\right) p(\beta) d\beta, \nonumber \\
  &=& \frac{1}{2}\int_{0}^{\infty} \mbox{erfc} \left(\sqrt{\beta}\right) \sum_{i=1}^{N_R} \left(-1\right)^{i-1} \, \bigg( \begin{array}{c} N_R \\ i  \end{array}\bigg) \nonumber \\
 &\times&
 \frac{2\,i}{2 i \bar{\gamma_0} - \bar{\gamma_1} }
\left(\exp \left[-\frac{\beta}{\bar{\gamma}_0} \right] - \exp\bigg[- \frac{2\,i \, \beta}{\bar{\gamma_1}} \bigg]\right)\,d\beta \nonumber \\
&=& \sum_{i=1}^{N_R} \left(-1\right)^{i-1} \, \bigg( \begin{array}{c} N_R  \\ i  \end{array}\bigg) \nonumber \\
 &\times&
 \frac{i}{2\,(2 i \bar{\gamma_0} - \bar{\gamma_1})} \bigg[\bar{\gamma_0} B_{z_0}[1,\frac{1}{2}] - \frac{\bar{\gamma_1}}{2\,i} B_{z_1}[1,\frac{1}{2}] \bigg], \nonumber \\
 &&
\label{BER_SCR}
\end{eqnarray}
where $z_0 = \frac{1}{\bar{\gamma_0}+1}$,  $z_1 = \frac{2\,i}{\bar{\gamma_1}+2\,i}$, and $B_x [a,b]$ is the incomplete beta function~\cite{TableofIntegrals}.

Note that $\gamma_1$ ($=\min(\gamma_{s,r^*}, \gamma_{r^*,d})$) sets
the upper-bound on the end-to-end (E2E) bit error rate (BER) of this
selection relaying scheme. However, the numerical examples discussed
below show that the performance evaluated using these derived
expressions are quite tight.

\subsection{Selection Relaying}
In this form of relaying it is assumed that the direct path is
unusable due to deep fade instances or heavy shadowing. Hence, the
BER performance can  be derived from expression given in
(\ref{BER_SCR}) by setting $\gamma_0 \rightarrow -\infty$. In
particular, after some manipulation, the following  error rate
expression can be obtained
\begin{eqnarray}
 BER_{sr} &=& \frac{1}{4} \sum_{i=1}^{N_R} \left(-1\right)^{i-1} \, \bigg( \begin{array}{c} N_R \\ i  \end{array}\bigg)   B_{z_1}[1,\frac{1}{2}],
\label{BER_SR}
\end{eqnarray}
which is a strikingly simple and compact expression.
%
\subsection{All-Path Selection Cooperating Relaying (ASR)}
In this relaying scheme the destination selects one path from the
$N_R+1$ possible paths ( the $N_R$ relay paths and the S-D path) for
signal detection. The scheme views the S-D link as a virtual relay
path (i.e., the S-R and R-D channels are the same). It then selects
a path according to the previous selection scheme. The important
distinction from the SCR is that the destination does not need to
perform maximal ratio combining. Therefore, a system using the ASR
scheme is  less complex than one using SCR. Although full diversity
order is obtained, the scheme is however, inferior to SCR in terms
of coding (or power) gain. The antenna gain advantage of SCR over
ASR is evident by comparing Figs.~\ref{FigBerSCR_SD} and
\ref{FigBERASR}. In these two figures, the corresponding curves have
the same slope, but the curves for SCR are shifted downward (i.e.,
in the direction of power gain or coding gain).

The ASR  combiner follows almost the same principle as
the traditional selection combining scheme (of collocated antennas)  with the distinction  that the broadcast nature of
 wireless channel is exploited and that the antennas (i.e., the relays) are distributed entities. The combiner in this case can be expressed
 as $\max\{ \min\{\gamma_{S-r^*}, \gamma_{r^*-D}\}, \gamma_{S-D}\}$, where $r^*$ is the best relay. The $\max \min $ formulation is essential to incorporate the fact that  the weak link constitutes the bottle-neck.

\section{Capacity and Outage Probability}
System capacity  and outage probability are information theoretic performance measures. Here, we demonstrate that the
analyses in this paper can be extended to calculating these performance measures.
The notion of capacity is valid where
 the channel is ergodic and there are no
constraints on the decoding delay on the receiver.  These conditions
are hardly met in practical communication systems. The channels do
behave in a manner such that there is no significant channel
variability.
Under such slow fading channel conditions, there is a non-zero error
probability that the channel will be in a deep fade. Therefore, it
is not possible to send a positive rate through the channel and yet
maintain
 a vanishingly small error probability, which explains why in strict sense, the  capacity of a
 slowly
 fading channel is zero. It is appropriate in this situation to consider outage probability. We note that outage and capacity
 are important communication modeling parameters, and we therefore provide expressions for evaluating them in the following discussion.
\subsubsection{Outage probability SCR}
An outage is defined as the event where the communication channel
does not support a target data rate under the SCR scheme, the outage
probability is given by the following expression (see the Appendix
for the derivation).
 \begin{eqnarray}
 p_{out,scr} &=&    \sum_{i=1}^{N_R} \left(-1\right)^{i-1} \, \bigg( \begin{array}{c} N_R  \\ i  \end{array}\bigg)
\bigg( 1 + \frac{1}{2\,i \bar{\gamma_0} - \bar{\gamma_1}}  \nonumber \\
 &\times&
\bigg[ \bar{\gamma_1}\exp \left( -  \frac{2 \, i \, a}{\bar{\gamma_1}} \right) - 2 i\,\bar{\gamma_0}\exp \left( - \frac{a}{\bar{\gamma}_0}  \right)\bigg] \bigg), \nonumber \\
&&
\label{outage2}
\end{eqnarray}
where $R$ is the target rate, and $ a = 2^{2 R} - 1.$
\subsubsection{Capacity of SCR}
The  ergodic channel capacity is considered. Therefore,  averaging
the instantaneous  channel capacity over the  fading distribution
has operational meaning. The capacity of the SCR scheme, in
bits/s/Hz, is given as
\begin{eqnarray}
\tilde{C} &=& \int_{0}^{\infty}\frac{1}{2} \log_2 (1 + \beta) p(\beta) d\beta \nonumber \\
 &=& \sum_{i=1}^{N_R}\left(-1\right)^{i-1} \,  \bigg( \begin{array}{c} N_R  \\ i  \end{array}\bigg) \nonumber \\
          &\times& \left(\frac{2}{(2\, \bar{\gamma_1} - 4 \,i \bar{\gamma_0})\ln 2} \right) \bigg[2\,i \bar{\gamma_0} \exp\left(\frac{1}{\bar{\gamma_0}}\right) \left(E_1 \bigg[ \frac{-1}{\bar{\gamma_0}}\bigg]  \right) \nonumber \\
          &-&  \bar{\gamma_1} \exp\left(\frac{2\,i}{\bar{\gamma_1}}\right) \left(E_1\bigg[\frac{-2\,i}{\bar{\gamma_1}}\bigg]   \right)    \bigg] \nonumber \\
           &=& \sum_{i=1}^{N_R}\left(-1\right)^{i-1} \, \left(\frac{1}{( \bar{\gamma_1} - 2 \,i \bar{\gamma_0})\ln 2} \right) \, \bigg( \begin{array}{c} N_R  \\ i  \end{array}\bigg) \nonumber \\
          &\times&  \bigg[ - 2\,i \bar{\gamma_0} \exp\left(\frac{1}{\bar{\gamma_0}}\right) E_1 \bigg[ \frac{1}{\bar{\gamma_0}}\bigg] 
          +  \bar{\gamma_1} \exp\left(\frac{2\,i}{\bar{\gamma_1}}\right)E_1\bigg[\frac{2\,i}{\bar{\gamma_1}}\bigg]    \bigg], \nonumber \\
          && 
\label{SCR_capacity}
\end{eqnarray}
where $E_1 [\cdot]$ is the exponential
integral~\cite{TableofIntegrals}. This capacity analysis also
generalizes the single relay treatment in~\cite{Mazen2005} to an
arbitrary number of relays. The derivation of this result, and those
below, is omitted due to space limitations.
\subsubsection{Outage Probability for SR}
The outage probability for the SR scheme can be expressed as,
\begin{eqnarray}
p_{out,sr} &=& \sum_{i=1}^{N_R} \left(-1\right)^{i-1}   \left(\begin{array}{c} N_R  \\ i  \end{array}\right)
 \nonumber \\
          & \times &
 \,\left(1 - \exp \left[ - \frac{2 i (2^{2\,R} -1)}{\bar{\gamma}}\right] \right).
\label{outage_sr}
\end{eqnarray}
\subsubsection{Capacity for the SR} The capacity of  this scheme is
\begin{eqnarray}
\tilde{C}& =& \sum_{i=1}^{N_R} \left(-1\right)^{i-1} \,  \left(\begin{array}{c} N_R  \\ i  \end{array}\right)
\frac{1}{\ln 2} \nonumber \\
&\times&
\exp \left(\frac{2 i}{\bar{\gamma}}\right) E_1\left( \frac{2\,i}{\bar{\gamma}} \right) \hspace{.25cm} \mbox{bits/s/Hz}.
\label{capacity}
\end{eqnarray}
%
\begin{figure}
\centering
 \includegraphics[scale=0.5]{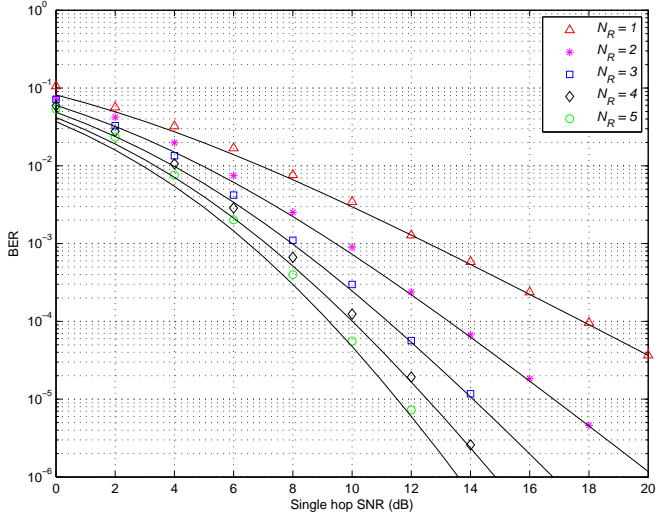}
 \vspace{-0cm}
  \caption{BER performance of the two-hop selection cooperative relaying scheme in Rayleigh fading. The S-D, R-D and S-R have the same
  average SNR.}\label{FigBerSCR_SD}
\end{figure}

\begin{figure}
\centering
 \includegraphics[scale=0.5]{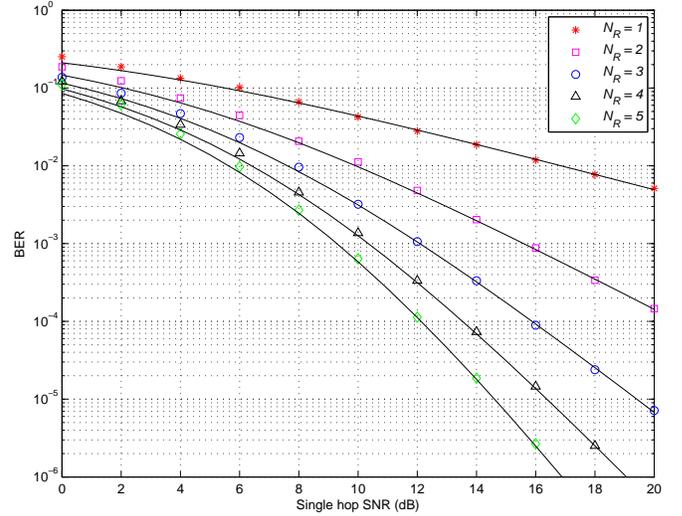}
 \vspace{-0cm}
  \caption{BER performance of the two-hop selection relaying scheme. The S-D link is not used.}\label{FigBER_SR}
\end{figure}

\begin{figure}
\centering
 \includegraphics[scale=0.5]{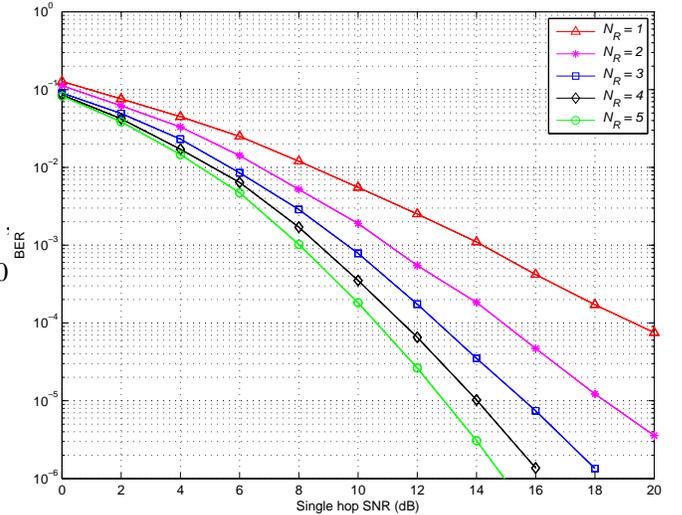}
 \vspace{-0cm}
  \caption{Simulated BER performance of the all-path selection  relaying scheme in Rayleigh fading.}\label{FigBERASR}
\end{figure}

\begin{figure}
\centering
 \includegraphics[scale=0.5]{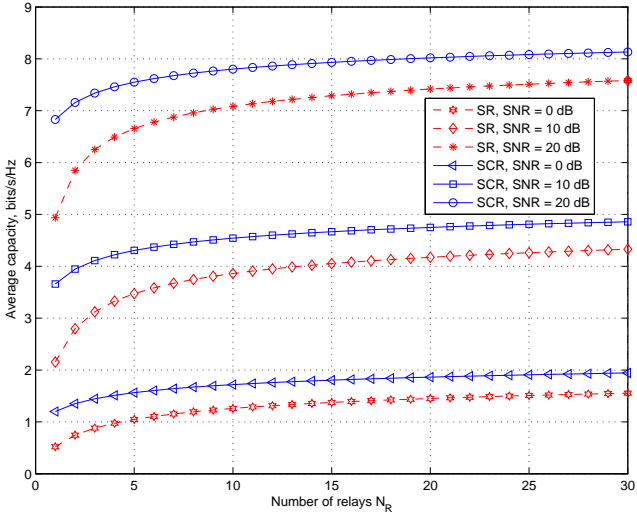}
 \vspace{-0cm}
  \caption{Average capacity of selection relaying and selection cooperative relaying in Rayleigh fading for different values of average SNR.}\label{FigCapacitySCRvsSR}
\end{figure}

\section{Numerical  Examples}
Figs.~\ref{FigBerSCR_SD},~\ref{FigBER_SR} and~\ref{FigBERASR} show
the bit error rate of selection cooperative relaying, selection
relaying and all path selection relaying schemes, respectively. BPSK
modulation is used on all the links, and slow Rayleigh fading is
assumed. In the SCR scheme, the S-D and S-R, and R-D links are
assumed to have the same average channel gains. All  receiving nodes
are assumed to have the same noise statistics. From
Figs.~\ref{FigBerSCR_SD} and~\ref{FigBER_SR}, simulation results
indicated with symbols, match closely with the analytical ones shown
as solid curves. From Fig.~\ref{FigBerSCR_SD}, it can be seen that a
diversity order equal to $N_R +1$ is obtained for an $N_R$ relay
network. This order of diversity can be calculated from the slope of
the curves. The same diversity order can be calculated from
Fig.~\ref{FigBERASR}. However, Fig.~\ref{FigBerSCR_SD} presents a
superior power gain advantage over Fig.~\ref{FigBERASR}.

The derived formulas for capacity are plotted in
Fig.~\ref{FigCapacitySCRvsSR}. The figure  compares the capacity of
the SCR and SR schemes. The capacity for SCR is shown as solid
curves and SR as dotted curves. The advantage of using the direct
path is also obvious from this figure, where a more than 11\%
increase in capacity is obtained over the capacity of the SR scheme.
A general observation is that the capacity saturates quickly with
the number of relays.

The analyses in this work can  be applied to systems that employ
coding and to systems that do not,  such as in sensor networks where
the node may have only detect-and-forward capability. In the latter
scenario, the error rate  is a valid performance criterion while in
the former, where coding is used (applicable to decode-and-forward
relaying) the outage or capacity is the reasonable performance
measure.

%
\section{Conclusion}
Much of the recent work on selection relaying has been focused on
information theoretic analyses. Bounds on high-SNR outage
probability have been presented in most of these studies.
 In contrast to earlier works, our contributions provide  error rate analyses that are reasonably accurate over
a large range of SNRs, most importantly in the low/medium SNR
region.  It is worth  noting  that practical systems operate at
considerably lower SNR values than the range where large SNR
analysis is accurate. It is important that network designers are
able to evaluate system performance with a reasonable degree of
accuracy to help them make meaningful decisions regarding network
design and system deployment.

\appendix
This section presents the derivation of the outage probability of
the selection cooperative relaying scheme.
\begin{eqnarray}
 p_{out,scr}   & =& \text{Pr}\, (I < R) \nonumber \\
         & =&  \text{Pr}\,\big(  \log \left(1 + \beta  \right) < 2\,R\big) \nonumber \\
 p_{out,scr}&=&   \int_{-\infty}^{a} \sum_{i=1}^{N_R} \left(-1\right)^{i-1} \, \bigg( \begin{array}{c} N_R  \\ i  \end{array}\bigg) \nonumber \\
&\times& \frac{2\,i}{2 i \bar{\gamma_0} - \bar{\gamma_1} }
\left(\exp \left[-\frac{\beta}{\bar{\gamma}_0} \right] - \exp\bigg[- \frac{2\,i \, \beta}{\bar{\gamma_1}} \bigg]\right) d\,\beta, \nonumber \\
&&
\label{out1}
\end{eqnarray}
where $a = 2^{2\,R} -1.$ Now by interchanging the integral and
summation operations, the integration can be performed, giving the
following:
\begin{eqnarray}
 &=&    \sum_{i=1}^{N_R} \left(-1\right)^{i-1} \, \bigg( \begin{array}{c} N_R  \\ i  \end{array}\bigg)
 \frac{2\,i}{2 i \bar{\gamma_0} - \bar{\gamma_1} } \nonumber \\
&\times& \int_{0}^{a} \left(\exp \left[-\frac{\beta}{\bar{\gamma}_0} \right] - \exp\bigg[- \frac{2\,i \, \beta}{\bar{\gamma_1}} \bigg]\right) d\,\beta, \nonumber \\
&=&    \sum_{i=1}^{N_R} \left(-1\right)^{i-1} \, \bigg( \begin{array}{c} N_R \\ i  \end{array}\bigg)
 \frac{2\,i}{2 i \bar{\gamma_0} - \bar{\gamma_1} }  \bigg( \gamma_0  - \frac{\bar{\gamma_1}}{2\,i} \nonumber \\
  &+& \frac{1}{2\,i} \left[\bar{\gamma_1} \exp\left[-\frac{2\,i a}{\bar{\gamma_1}}\right] - 2 i \bar{\gamma_0} \exp\left[\frac{-a}{\bar{\gamma_0}}\right] \right]\bigg). \nonumber \\
\mbox{Finally,} \\
p_{out,SCR}&=&    \sum_{i=1}^{N_R} \left(-1\right)^{i-1} \, \bigg( \begin{array}{c} N_R \\ i  \end{array}\bigg)
  \bigg( 1 \nonumber \\
 &+&   \frac{1}{2 \,i \bar{\gamma_0} - \bar{\gamma_1} }
 \left[\bar{\gamma_1} \exp\left(-\frac{2\,i a}{\bar{\gamma_1}}\right) - 2 i \bar{\gamma_0} \exp\left(\frac{-a}{\bar{\gamma_0}}\right) \right]\bigg). \nonumber \\
 &&
\label{out2}
 \end{eqnarray}
This completes the derivation.
The outage performance expression for the selection relaying scheme (without the direct path)  can be derived  in a similar way through the PDF in (\ref{pdf_SR}). However,
it can also be obtained from~(\ref{out2}). The expression  is given as
\begin{eqnarray}
        p_{out,sr} & =&  \int_{-\infty}^{2^{2R} -1 } p(\gamma) \,d\gamma \nonumber \\
        &=&    \sum_{i=1}^{N_R} \left(-1\right)^{i-1} \, \bigg( \begin{array}{c} N_R  \\ i  \end{array}\bigg)
 \left( 1   -
 \exp\left[-\frac{2\,i a}{\bar{\gamma}}\right] \right). \nonumber \\
 &&
\label{out_sr}
 \end{eqnarray}

 \end{document}